\documentclass[reprint,
superscriptaddress,
amsmath,amssymb,
aps,prl,
twocolumn,
showpacs,
]{revtex4-1}
\usepackage{tikz}
\usepackage{graphicx}
\usepackage{dcolumn}
\usepackage{bm}
\usepackage{braket}
\usepackage{hyperref}
\usepackage{amsmath}
\usepackage{graphicx}
\usepackage{fancyhdr}
\usepackage{float}
\usepackage{todonotes}
\usepackage{xcolor}
\usepackage{xspace}
\usepackage{eucal}
\usepackage{ulem}
\usepackage{todonotes}
\usepackage{multirow}
\usepackage[utf8]{inputenc}
\usepackage{pgfplots}
\pgfplotsset{compat=1.14}
\usepackage{upgreek}
\usepackage{siunitx}
\usepackage{gensymb}
\usepackage{amsmath}
\usepackage{caption}
\usepackage{subcaption}
\usepackage{textcomp}
\usepackage{comment}
\usepackage{siunitx}
\usepackage[justification=justified,singlelinecheck=false]{caption}
\usepackage{todonotes}
\usepackage{cleveref}

\Crefname{figure}{Figure}{Figures}
\usepackage{siunitx}
\DeclareSIUnit\g{g}
\DeclareSIUnit\gal{Gal}
\DeclareSIUnit\torr{Torr}
\DeclareSIUnit\inch{inch}
\DeclareSIUnit\joule{J}
\begin{document}
\title{Optomechanical Inertial Sensors}
\author{Adam Hines}
\author{Logan Richardson}
\author{Hayden Wisniewski}
\author{Felipe~Guzman}\email[Electronic mail: ]{felipe@optics.arizona.edu}
\affiliation{James C. Wyant College of Optical Sciences, University of Arizona, 1630 E. University Blvd., Tucson, AZ 85721, USA}%
\date{\today}
%
\begin{abstract}
We present a performance analysis of compact monolithic optomechanical inertial sensors that describes their key fundamental limits and overall acceleration noise floor. Performance simulations for low frequency gravity-sensitive inertial sensors show attainable acceleration noise floors of the order of  $\SI{1e-11}{\meter\per\second\squared\per\sqrt{\hertz}}$. Furthermore, from our performance models, we determine the optimization criteria in our sensor designs, sensitivity, and bandwidth trade space. We conducted characterization measurements of these compact mechanical resonators, demonstrating $mQ$-products at levels of \SI{250}{kg}, which highlight their exquisite acceleration sensitivity.
\end{abstract}
\maketitle
\section{Introduction}
Commercially available high-sensitivity inertial sensors are typically massive systems that are not easily transportable and deployable due to their total mass and dimensions. Conversely, compact commercial systems, while easily transportable and field capable, exhibit comparatively higher acceleration noise floors.

Spring gravimeters and relative gravimeter technologies \cite{zhili,4178879,HAO2009196} tend to be large, expensive, and offer limited sensitivity. These systems use a mass-spring system, which measures the local gravitational acceleration by tracking the spring extension \cite{springgravi,Weber1966}, usually with electrostatic measurement techniques. One such example is the Scintrex CG-6 gravimeter that can achieve acceleration sensitivities of $\SI{e-9}{\g\per\sqrt{\hertz}}$ over a bandwidth of up to \SI{10}{\hertz} \cite{CG6}. Superconducting relative gravimeters create ideal springs by levitating a superconducting niobium sphere in a non-uniform magnetic field \cite{Goodkind1999,Prothero1972}. In this way, one can measure the local gravity with a sensitivity of $\SI{e-9}{\m\per\second\squared\sqrt{\hertz}}$ over a bandwidth of \SI{250}{\milli\hertz} \cite{iGrav}. However, due to the intensive operation requirements and maintenance, these systems are not suitable for deployment, since exposure to large accelerations, as usual on the field, can cause tares in the data. Micro-electromechanical systems (MEMS) are typically small and low-cost in comparison to other types of gravimeters and utilize small mass-spring systems that are read out electrostatically. Recent development in MEMS devices have demonstrated sensitivities at levels of $\SI{30}{\nano\g\per\sqrt{\hertz}}$ over a bandwidth of \SI{1}{\hertz} \cite{Middlemiss2016,8589884,mi7090167}, however, these sensitivity levels are comparatively lower by one to two orders of magnitude with respect to other commercial systems.

Absolute gravimeters, such as the Micro-G Lacoste FG5 and atom interferometers, offer long term stability in gravitational measurements \cite{FG5,AtomInterferometer}. When operated alone, however, they are susceptible to external vibrations, which obscure the acceleration measurement and ultimately limit the performance and their deployment capabilities for field operation \cite{FG5}. Furthermore, the Micro-G Lacoste FG5 require cost-intensive and frequent calibrations, and the aging of the springs causes drift over time. Atom interferometry, on othe other hand, is a technology still under intensive development.

Advances in optomechanics over the past decade and research into their fundamental limits have paved the way for the development of novel compact and highly sensitive inertial sensors. The thermal acceleration noise floor and mechanical losses have been studied extensively, for example, in the context of suspensions and mechanical systems for ground-based gravitational wave observatories, such as the Laser Interferometer Gravitational-Wave Observatory (LIGO)\cite{Cumming2009,Cumming2012}.

In this article, we present the results of our investigations regarding compact optomechanical inertial sensors that consist of monolithically micro-fabricated fused silica mechanical resonators and experimentally demonstrate high acceleration sensitivities and measure $mQ$-products above \SI{240}{\kilo\gram}. Furthermore, we studied the mechanics of our compact mechanical resonators using computational simulations, particularly on various loss mechanisms that would impact their sensitivity and conducted a trade-off analysis to determine resonator topologies that exhibit the best performance. These results guide our efforts developing novel compact and highly sensitive optomechanical inertial sensors.

Our optomechanical sensors provide numerous advantages over traditional acceleration sensing technologies due to their comparatively compact size and low mass, as well as their inherent vacuum compatibility, optical readout, and monolithic composition. Here, we present an optomechanical resonator, capable of achieving acceleration noise floors at levels of $\SI{1e-11}{\meter\per\second\squared\per\sqrt{\hertz}}$ with a footprint of \SI{48}{\milli\meter} $\times$ \SI{92}{\milli\meter} and a mass of \SI{26}{\gram}, making it small and transportable. The optical laser-interferometric readout of our sensor provides a significantly higher sensitivity than typical electrostatic techniques, and is insensitive to external electro-magnetic fields. Moreover, our sensors are monolithically fabricated from very low loss materials, such as fused silica, allowing us to achieve high mechanical quality factors. 

Similar compact, monolithic, optomechanical sensors with high resonant frequencies have already shown excellent acceleration noise floors at the nano-g \SI{}{\per\sqrt{\hertz}} over \SI{10}{\kilo\hertz}, as well as laser-interferometric displacement sensitivites of $\SI{1e-16}{\meter\per\sqrt{\hertz}} $\cite{Guzman2014}.

In this article we present a prototype optomechanical sensor with a resonant frequency of \SI{10}{\hertz} that targets high acceleration sensitivities at low frequencies. Lowering the resonant frequency of the sensor increases the acceleration sensitivity. The portability, comparatively low cost, and monolithic composition of our devices make them excellent candidates for a broad spectrum of applications, including gravimetry and gravity gradiometry, geodesy, seismology, inertial navigation, vibration sensing, metrology, as well as other applications in geophysics. The sensor design and performance modelling presented in this paper allow us to understand how we can use optomechanical sensors as low frequency accelerometers, and to understand their sensitivity limits.

\section{ sensitivity and losses in optomechanical inertial sensors}
\begin{figure}[htbp]
\centering
\fbox{\includegraphics[width=.95\linewidth]{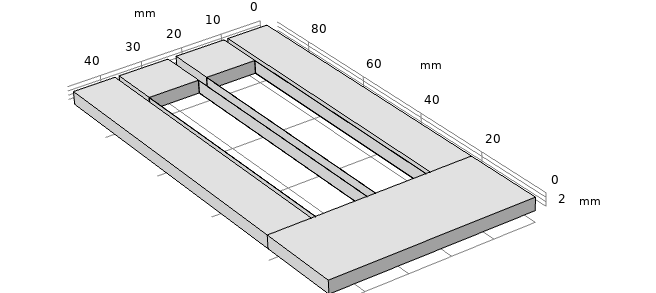}}
\caption{Geometry of our optomechanical resonator generated in COMSOL. From this model, we calculate the mechanical properties of our sensor, including the resonant frequency and energy loss mechanisms. The sensor described in this work has overall dimensions of $\SI{48}{\milli\meter} \times \SI{92}{\milli\meter} \times \SI{3}{\milli\meter}$ and mass of \SI{26}{\gram}. Two oscillating test masses each with a mass of \SI{0.95}{\gram} are supported by two flexures each with a thickness of \SI{100}{\micro\meter}. The wafer has two oscillators with the intent to increase $Q$ via coupled motion in the individual test masses.}
\label{fig:COMSOL-pic}
\end{figure}

Our sensor consists of monolithic fused silica resonators based on a parallelogram dual flexure design that supports the oscillating acceleration-sensitive test mass (Figure~\ref{fig:COMSOL-pic}). We use a displacement readout laser interferometer to measure the dynamics of the test masses (Section~\ref{sec:displacement_interferometry}). The total mass of our resonator head is approximately \SI{26}{\gram} with an oscillating test mass of \SI{0.95}{\gram}. The spring flexures supporting the test mass are \SI{0.1}{\milli\meter} thick by \SI{60}{\milli\meter} long which yields a resonance frequency of \SI{10}{\hertz}. We place an aluminum-coated fused silica mirror on top of the test mass with no adhesive for the laser interferometer, which reduces the resonant frequency to \SI{3.76}{\hertz} due to the added \SI{1.25}{\gram} mass. Micro-fabricated by laser-assisted dry-etching, the overall resonator is \SI{48}{\milli\meter} x \SI{92}{\milli\meter} x \SI{3}{\mm}, making it very compact, and is constructed from a monolithic fused silica wafer for its low-loss properties at room temperature \cite{Ageev2004}, which makes these devices easy to operate and deploy in the field. Low losses in fused silica result in low frequency-dependent damping, high mechanical quality factors, and low thermal noise.

The acceleration sensitivity of the optomechanical sensors is limited by the thermal noise floor of its oscillating test mass and the sensitivity of the test mass displacement sensor. 
Within an optomechanical resonator, there are various  mechanisms that dissipate energy. These mechanisms can be separated into two categories: external velocity damping (\emph{eg.} gas damping) and internal damping (\emph{eg.} surface damage). Therefore, we treat the resonator as a mass-spring system with a velocity damping term and a complex spring constant. The equation of motion of such a system is given by \cite{Saulson1990}:
\begin{equation}
    F = m\ddot{x} + m\Gamma_{v}\dot{x} + m\omega_{0}^{2}(1+i\phi(\omega))x
    \label{eqn_of_motion},
\end{equation}
where $m$ is the resonator's mass, $\Gamma_{v}$ is the velocity damping rate, $\omega_{0}$ is the resonant frequency, and $\phi(\omega)$ is the loss coefficient for internal losses. The thermal motion of the resonator is derived from Equation~\ref{eqn_of_motion} via the Fluctuation-Dissipation Theorem \cite{Saulson1990}. Using this technique, we find the power spectral density of the thermal motion to be:
\begin{equation}
    x_{th}^{2}(\omega) = \frac{4k_{B}T}{m\omega}\frac{\omega\Gamma_{v}+\omega_{0}^{2}\phi(\omega)}{(\omega_{0}^{2} - \omega^{2})^{2} + (\omega\Gamma_{v}+\omega_{0}^{2}\phi(\omega))^{2}},
    \label{disp-psd}
\end{equation}
where $k_{B}$ is Boltzmann's constant, $T$ is temperature, and $\omega$ is angular frequency. Furthermore, from Equation~\ref{eqn_of_motion} we also note that the transfer function relating displacement to acceleration is given by:
\begin{equation}
    \frac{x(\omega)}{a(\omega)} = \frac{-1}{\omega_{0}^{2} - \omega^{2} + i(\omega\Gamma_{v}+\omega_{0}^{2})}.
    \label{transfer function}
\end{equation}
Using Equation~\ref{disp-psd} and Equation~\ref{transfer function}, we immediately find that the thermal acceleration noise is given by:
\begin{equation}
    a_{th}^{2}(\omega) = \frac{4k_{B}T}{m\omega}(\omega\Gamma_{v}+\omega_{0}^{2}\phi(\omega)).
    \label{acc-psd}
\end{equation}
By inspection, we see that in the high-frequency regime ($\omega\Gamma_{v} >> \omega_{0}^{2}\phi(\omega)$), Equation~\ref{acc-psd} is dominated by velocity damping and asymptotically approaches a constant. This is consistent with observations of uniform thermal noise at high frequencies. At low frequencies, however, the thermal acceleration noise is dominated by internal losses and has a $\omega^{-1}$ dependence.

In order to predict the thermal motion of our optomechanical resonator, we need to know the velocity damping rate and the mechanical loss coefficient. The velocity damping is dominated by gas damping, which is determined computationally and is discussed further in Section~\ref{sec:computational_analysis}\ref{sec:gas_damping}. The mechanical loss coefficient originates from four main loss mechanisms in our resonators: bulk losses, surface losses, thermoelastic losses, and anchor losses. Thus, the total mechanical loss in the resonator is given by \cite{Cumming2009,Cumming2012}:
\begin{equation}
\begin{aligned}
    \phi(\omega) = \phi_{surface}+\phi_{bulk}(\omega)+\phi_{thermo}(\omega)\\
    +\phi_{anchor}(\omega) \label{totalLoss}.
\end{aligned}
\end{equation}

In our analysis, we assume negligible variation in flexure thickness, which was experimentally verified by microscope measurements to be less than \SI{10}{\micro\meter} along the full length. Furthermore, since elastic energy is stored in the flexures, we do not consider the test mass geometry in our analysis. We further discuss this point in Section~\ref{sec:sensor_performance}.

\subsection{Bulk and Surface Losses}
Bulk losses are a result of energy losses intrinsic to the material. We use the model experimentally determined by Penn \emph{et al.} \cite{PENN20063} to determine the contribution from bulk losses in our resonator:
\begin{equation}
    \phi_{bulk}(\omega) = 7.6\times10^{-10}\bigg(\frac{\omega}{2\pi}\bigg)^{0.77}. \label{phiBulk}
\end{equation}
Surface losses encapsulate the intrinsic losses at the surface of the material resulting from damage, or surface imperfections from manufacturing process. We can model surface losses for an arbitrary flexure shape as\cite{GRETARSSON2000108}:
\begin{equation}
    \phi_{surface} = \mu h\phi_{s}\frac{S}{V} \label{phiSurf},
\end{equation}
where $\mu$ is a constant dependent on the shape of the flexure, $h$ is the skin-depth of the surface, $\phi_{s}$ is the intrinsic loss at the surface, $S$ is the surface area of the flexure, and $V$ is its volume. The fabrication of our resonator could lead to a substantial amount of surface losses if the surface quality is non-ideal. 
However, the surface losses for ideal fused silica is better documented than that for non-pristine samples. To determine the surface losses for ideal flexures in a given geometry, we model the surface losses of ideal fused silica fibers; such as those which are flame or laser-pulled and exhibit high surface quality. For such fibers, Gretarsson \emph{et al.} experimentally determined $h\phi_{s}$ to be \SI{6.15}{\pico\meter}. For a flexure with a rectangular cross section, Equation~\ref{phiSurf} becomes:
\begin{equation}
    \phi_{surface} = \frac{3+A}{1+A}h\phi_{s}\frac{2(x+y)}{xy} \label{phiSurf2},
\end{equation}
where $x$ and $y$ are the flexureâ€™s cross-section dimensions, and $A$ is the aspect ratio of the rectangular cross-section \cite{GRETARSSON2000108}. Our flexures have \SI{0.1}{\milli\meter} $\times$ \SI{3}{\milli\meter} cross-section and an aspect ratio of 30. 

\subsection{Thermoelastic Losses}
Thermoelastic losses describe bending of the flexures due to spontaneous temperature fluctuations and can be theoretically derived from:
\begin{equation}
    \phi_{thermo}(\omega) = \frac{YT\alpha^{2}}{\rho C}\frac{\omega\tau}{1+\omega^{2}\tau^{2}} \label{phiTherm},
\end{equation}
where $Y$ is the Young's modulus, $\rho$ is the mass density, $C$ is the specific heat capacity, and $\alpha$ is the coefficient of thermal expansion \cite{Cumming2009,Cumming2012}. For fused silica, these values are $Y=\SI{71.5}{\giga\pascal}$, $\rho=\SI{2203}{\kilo\gram\per\meter\cubed}$, $C=\SI{670}{\joule\per\kilo\gram}$, and $\alpha = \SI{5.5e-7}{\per\kelvin}$. In our simulations we assume operation at room temperature, $T=\SI{293}{\kelvin}$. The term $\tau$ is the characteristic time needed for heat to travel across the cross section of the flexure. For rectangular cross sections, this time is given by:
\begin{equation}
    \tau = \frac{\rho Ct^{2}}{\pi^{2}\kappa} \label{time},
\end{equation}
where $t$ is the thickness of the flexure and $\kappa$ is the thermal conductivity \cite{Cumming2009,Cumming2012}. The thermal conductivity of fused silica is taken to be $\kappa=\SI{1.4}{\watt\per\meter\per\kelvin}$.

\section{Optomechanical Inertial Sensor Performance}
\label{sec:sensor_performance}
\begin{figure}[htbp]
\centering
\fbox{\includegraphics[width=.95\linewidth]{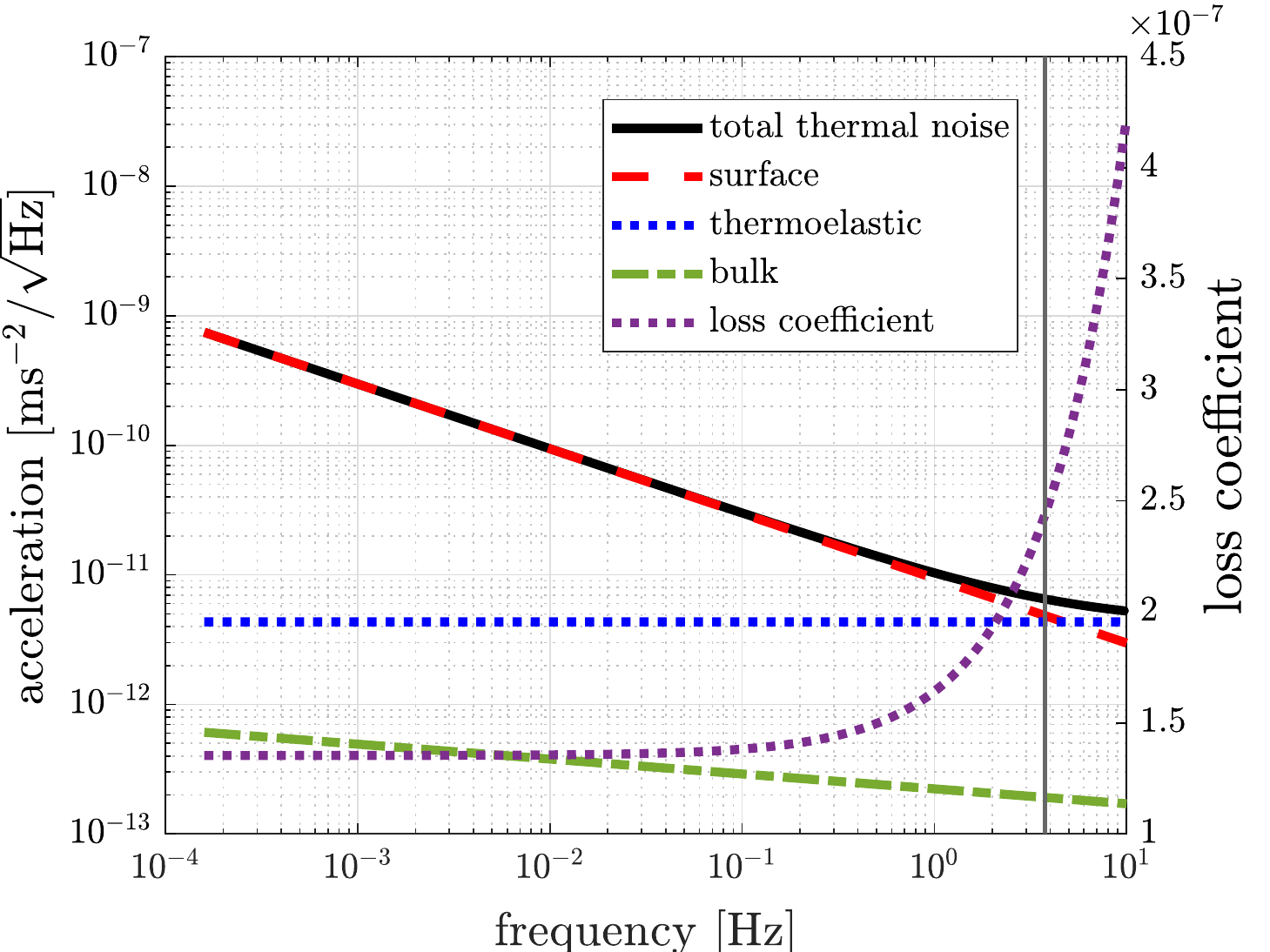}}
\caption{The calculated linear spectral density of acceleration thermal noise for a \SI{3.76}{\hertz}, \SI{2.2}{\gram} test mass is plotted on the left axis. The resonant frequency is denoted by a vertical line. We also plotted the contribution from each loss mechanism, from which we can see that surface losses are the dominant noise source for frequencies below resonance. On the right axis, the loss coefficient is plotted as a function of frequency.}
\label{fig:acc-noise}
\end{figure}

\begin{figure}[htbp]
\centering
\fbox{\includegraphics[width=.95\linewidth]{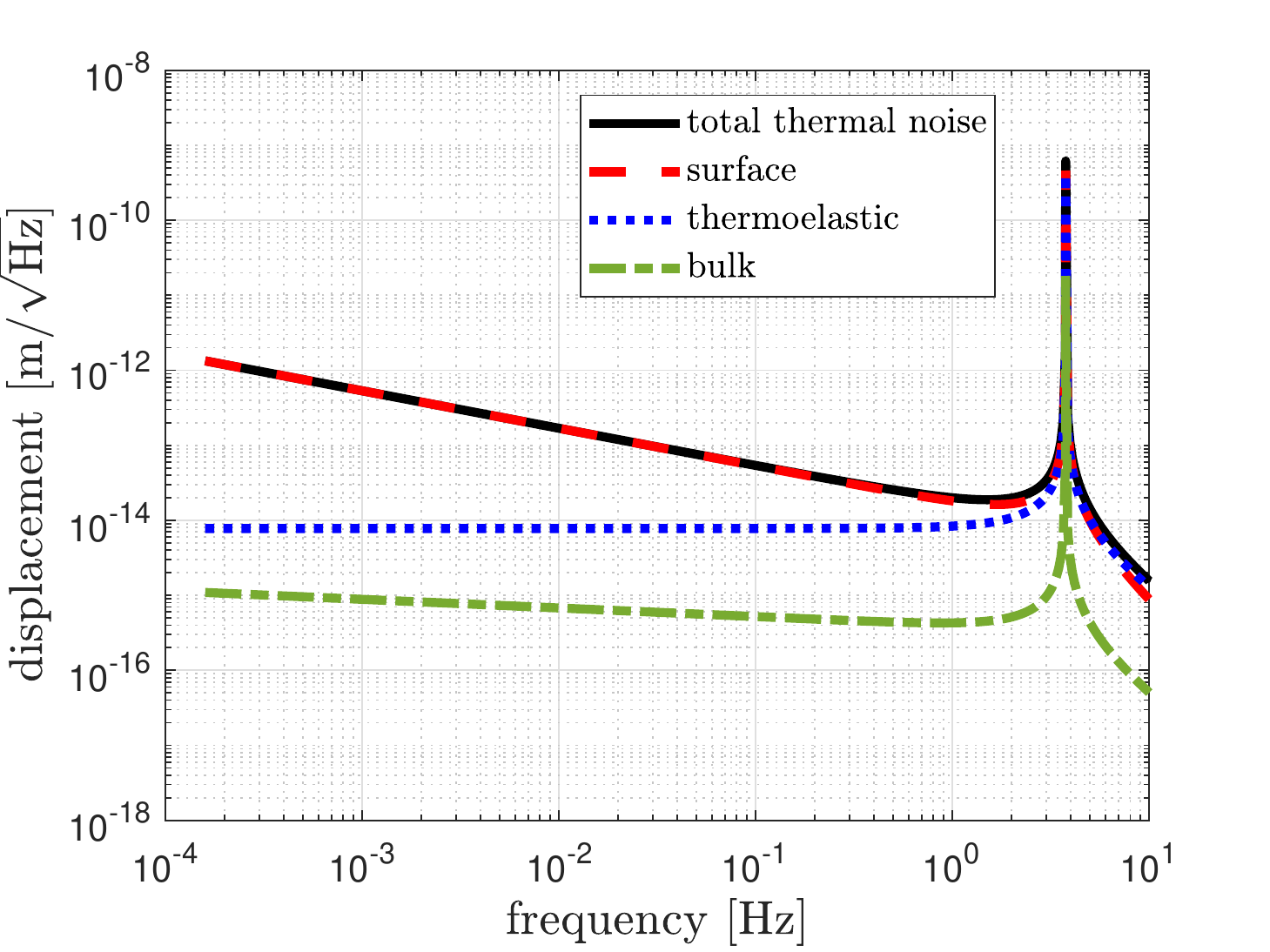}}
\caption{Calculated linear spectral density of displacement thermal noise floor for a \SI{3.76}{\hertz}, \SI{2.2}{\gram} test mass. Each loss mechanism's contribution is included. As shown, a read-out system would need to have a displacement sensitivity on the order of \SI{1e-13}{\m\per\sqrt{\hertz}} to resolve the thermal noise floor.}
\label{fig:disp-noise}
\end{figure}

From Equation~\ref{disp-psd} and Equation~\ref{acc-psd}, we can calculate the individual contributions from each loss mechanism to compute the expected displacement and acceleration amplitude spectral densities for a given resonator. Assuming operation at sufficiently low pressures for the velocity damping rate to be negligible with respect to other loss mechanisms, we show the computed linear spectral densities in Figures~\ref{fig:acc-noise} and \ref{fig:disp-noise}. For an oscillator with a resonant frequency of \SI{3.76}{\hertz} and a test mass $m =\SI{2.2}{\gram}$, surface losses dominate the spectrum at low frequencies with thermoelastic losses only becoming relevant near resonance. We observe that bulk losses are a much smaller contribution compared to the other mechanisms for the bandwidth of interest. This is consistent with Equations~\ref{phiBulk}, \ref{phiSurf}, and \ref{phiTherm}, which suggest that the bulk losses are several orders of magnitude smaller than those from the other loss mechanisms.

\section{Computational analysis}
\label{sec:computational_analysis}
\subsection{Simulated gas damping}
\label{sec:gas_damping}

\begin{figure}[htbp]
\centering
\fbox{\includegraphics[width=.95\linewidth]{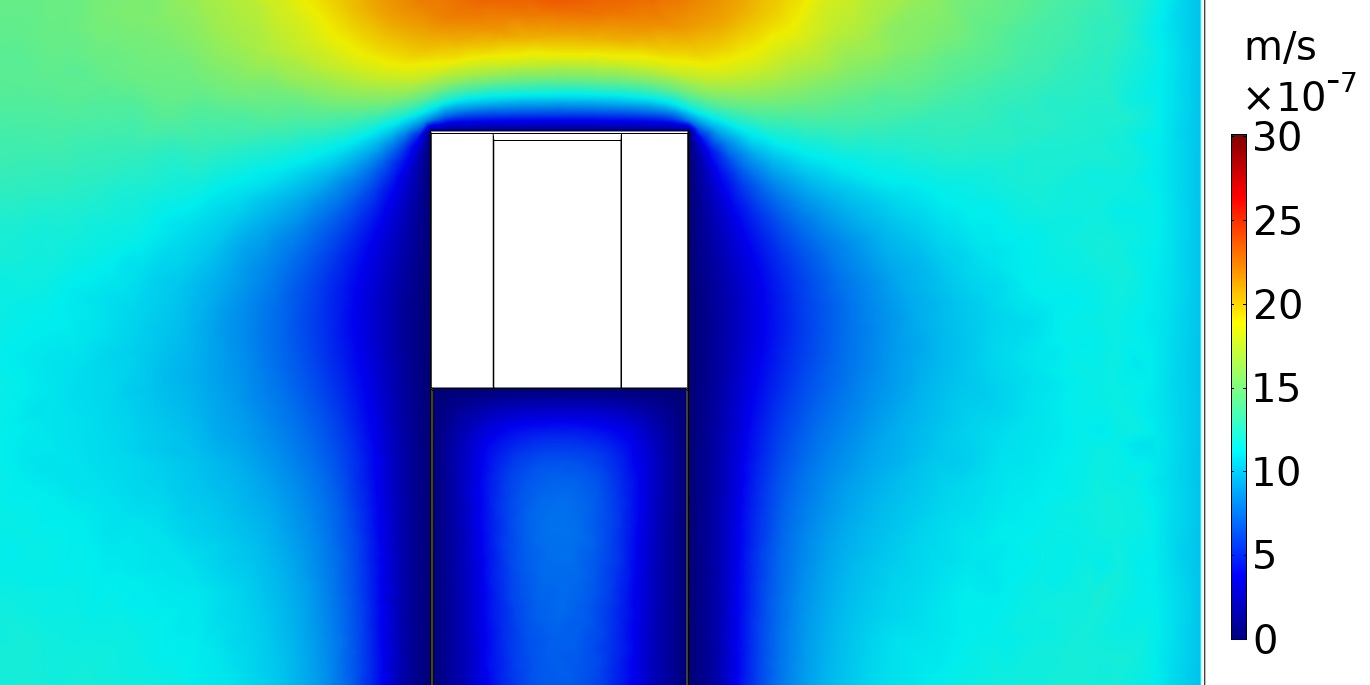}}
\caption{Simulated airflow around an optomechanical resonator. Lighter colors indicate higher airflow speed, whereas darker colors indicate a lower velocity. The inlet of the airflow is the left of the test mass, causing the air to move to the right. By integrating the pressure along the surface of the test mass and flexure, we compute the air drag force and mechanical quality factor of the resonator at atmospheric pressures.}
\label{fig:airflow}
\end{figure}

To better understand how the flexure and test mass geometry affect loss mechanisms, we utilized finite element analysis and modeled our resonator in COMSOL 5.4 (see Figure~\ref{fig:COMSOL-pic}). We used the Solid Mechanics module to calculate the eigenfrequencies of the resonator, and the Creeping Flow fluid dynamics module to estimate the quality factor of this resonator at atmospheric pressure. We modeled the mechanical oscillator assuming a large airbox surrounding the test mass and its flexures. The inlet and outlet of the air box were given a pressure differential that generated a \SI{1}{\micro\meter\per\second} air current at the test mass position. We used COMSOL to calculate the steady-state solution to this airflow, and we found the net force acting on the test mass by integrating the pressure over its surface (Figure~\ref{fig:airflow}).  From the calculated force, we found the linear drag coefficient, and then we applied the air resistance to the test mass as a boundary load. Performing an eigenfrequency analysis, this time without the air box, COMSOL produced the mechanical quality factor of the resonator in air. For our \SI{10}{\hertz} resonator, we found this $Q$ to be $\approx 700$ with a damping rate of \SI{8.97e-2}{\per\second}.

\subsection{Simulated elastic energy distribution}
\begin{figure}[htbp]
\centering
\fbox{\includegraphics[width=.95\linewidth]{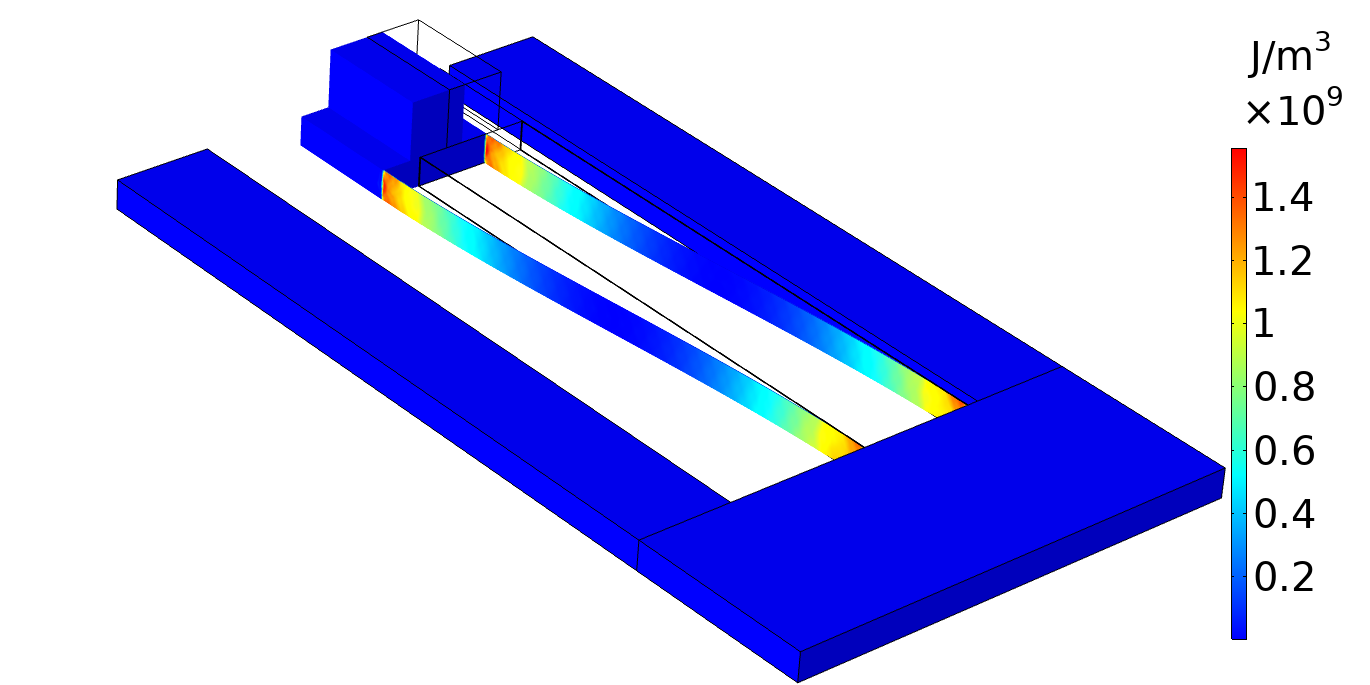}}
\caption{A COMSOL simulation of the elastic energy density in our inertial resonator at its resonant frequency with a mirror on top of the test mass. The units on the legend are arbitrary. In this figure, red represents a greater energy density, and blue indicates a low energy density. The outline represents the equilibrium position of the test mass. From this simulation, we confirm that mechanical losses in our resonator are mostly located within the flexures, as opposed to within the test mass. }
\label{fig:elastic_energy}
\end{figure}

To support the claim that we only need to consider the flexure geometry to calculate the mechanical losses, we use COMSOL to calculate the elastic energy density throughout the resonator. When performing an eigenfrequency analysis of the resonator, COMSOL outputs the distribution of elastic energy. From this, we calculate that 99.77\% of the elastic energy is located within the flexures, 0.143\% is within the test mass, and 0.087\% is within the remainder of the fused silica wafer. We depict the elastic energy density in Figure~\ref{fig:elastic_energy}. Evaluation of the bulk, surface, and thermoelastic losses for the test mass yields a mechanical loss coefficient of \num{2.0e-7}. However, when weighted by the amount of elastic energy stored in the test mass, this gives a net mechanical loss coefficient of \num{2.91e-10}. This value is negligible since it is more than three orders of magnitude lower than the losses in the flexures $(\approx\num{4.8e-7})$. Furthermore, the fraction of the energy stored in the mirror on the test mass was found to be \num{6.9e-8}, suggesting that the losses from the mirror are negligible as well.

\begin{figure}[htbp]
\centering
\fbox{\includegraphics[width=.95\linewidth]{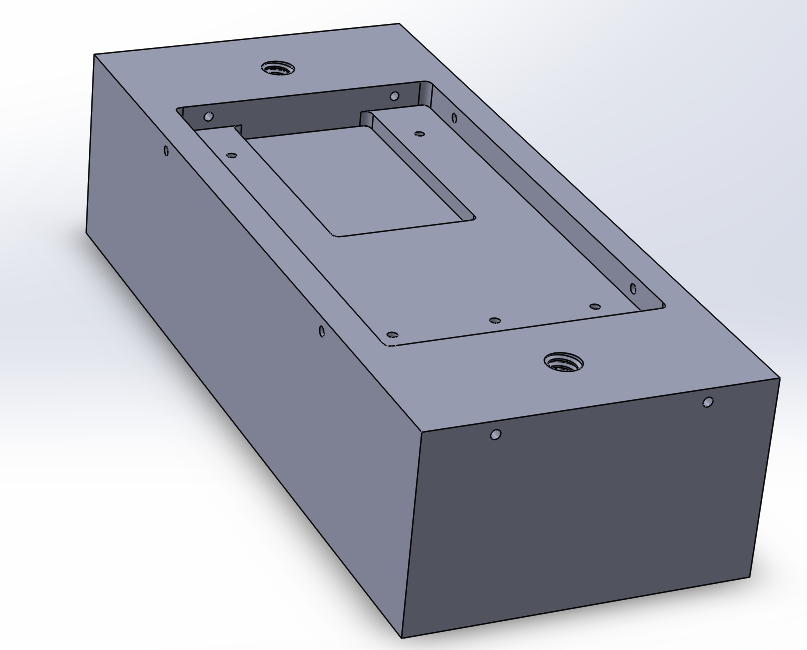}}
\caption{Model of resonator mount. The small holes in the walls and bottom of the mount hold ball bearings, which prevent the fused silica from contacting the aluminum mount. The larger holes are threaded to place a lid over the resonator. This mount also secures the resonator so that it can measure in the vertical orientation.}
\label{fig:mount}
\end{figure}

\subsection{Simulated anchor losses}
\label{sec:anchor_losses}
We can extend this exercise to estimate anchor losses by modeling the mounting apparatus. To mitigate anchor losses, and for the purpose of testing and characterizing our optomechanical inertial sensor, we fabricated a mount for the resonator that reduces the contact area between the fused silica and the rough aluminum surface of the mount with higher losses. This mount holds the resonator in place using thirteen \SI{3/32}{\inch} diameter aluminum ball bearings, which limits the amount of energy lost to the mounting apparatus, and allows us to tilt the sensor vertically. Figure~\ref{fig:mount} depicts a computer rendering of this mount. The eigenfrequency analysis tells us that the fraction of the elastic energy contained within the ball bearings is \num{3.4e-7}. The internal losses of aluminum are on the order of \num{1e-3} \cite{AlLoss}. Therefore, we can safely expect the losses from the mounting apparatus to be several orders of magnitude lower than the losses from the flexures, meaning that anchor losses are not the dominant loss mechanism in our current setup.

\subsection{Optimization of flexure dimensions}
\begin{figure}[htbp]
\centering
\fbox{\includegraphics[width=.95\linewidth]{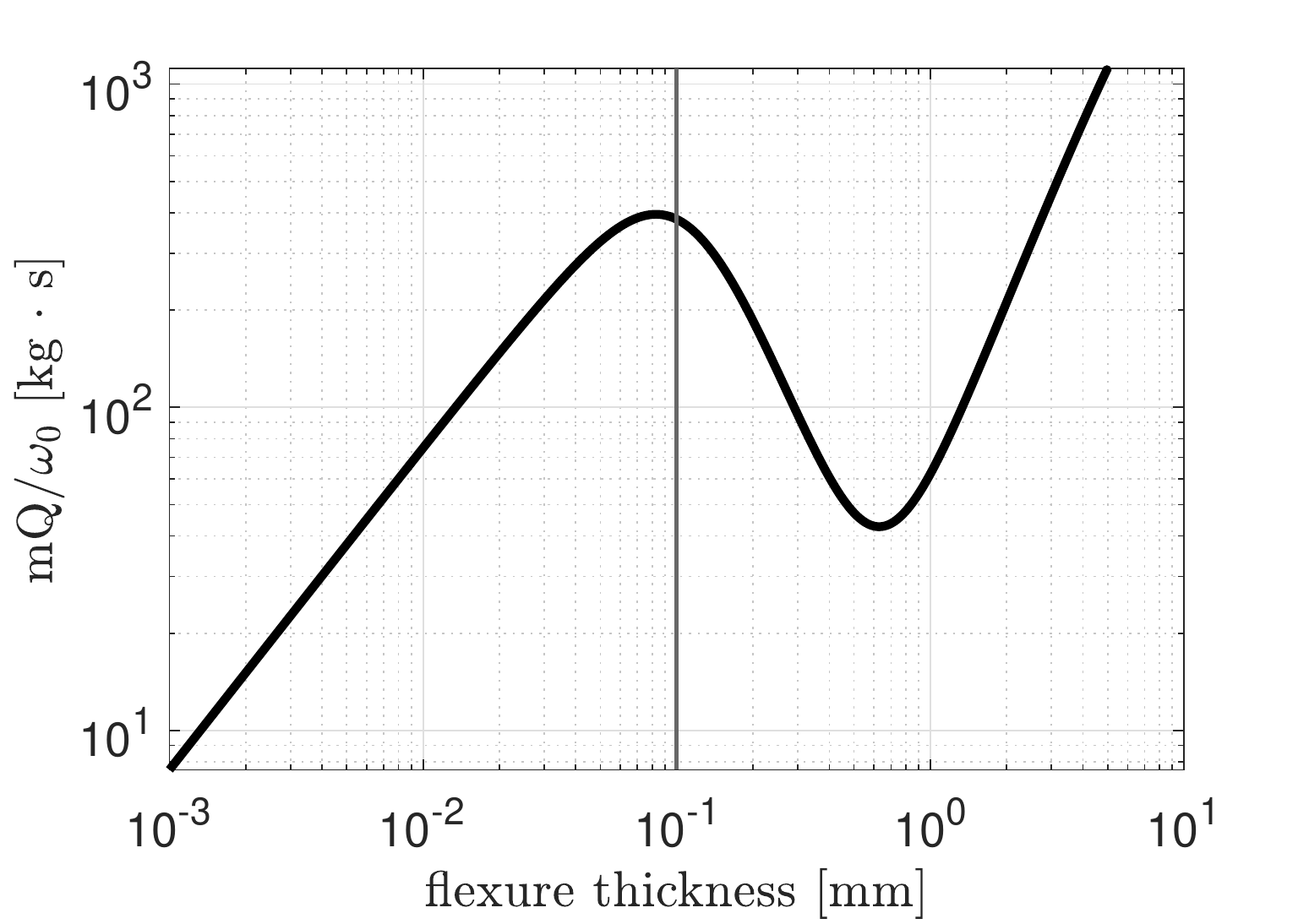}}
\caption{An optimization model of $mQ/\omega_{0}$ for various flexure thicknesses of a \SI{3.76}{\hertz} resonator. In this simulation, the resonant frequency is held constant. When varying the thickness of the flexures' smallest dimension, we assume the length of the flexures also vary to keep the resonance constant. We evaluated the surface, bulk, and thermoelastic loss models for a range of flexure thicknesses. The local maximum around \SI{8.3e-2}{\milli\meter} indicates the optimum flexure thickness which yields the lowest noise floor. For comparison, our current resonator has \SI{0.1}{\milli\meter} flexures, denoted by a vertical line in the plot.}
\label{fig:optimization}
\end{figure}

In addition to calculating the mechanical quality factor, Equation~\ref{totalLoss} allows us to compute the flexure dimensions that optimize the resonator sensitivity. By noting that the mechanical quality factor of a resonator is related to the loss coefficient by:
\begin{equation}
    Q = \frac{1}{\phi(\omega_{0})}
    \label{Q-phi},
\end{equation}
we evaluate Equations~\ref{totalLoss}, \ref{phiBulk}, \ref{phiSurf}, and \ref{phiTherm} for a given resonance and range of flexure thicknesses. In doing so, we determine the quality factor as a function of the flexure dimensions. We then optimize the dimensions by finding the thickness and length combination that produces the desired resonance with the largest $mQ/\omega$ value, ensuring that the optimized dimensions produce the lowest acceleration noise floor possible. The optimization of a \SI{3.76}{\hertz}, \SI{2.2}{\gram} test mass is depicted in Figure~\ref{fig:optimization}. We see that there is a local $mQ/\omega_{0}$ maximum when the flexure thickness is approximately $t=\SI{0.083}{\milli\meter}$. At this thickness, $mQ/\omega_{0} = \SI{392}{\kilo\gram\cdot\second}$ and $Q\approx\num{4.2e6}$. Such a resonator would have a thermal noise floor of approximately $\SI{1.0e-11}{\meter\per\second\squared\per\sqrt{\hertz}}$. In principle, we can potentially achieve even larger $mQ/\omega_{0}$-values for thicker (>\SI{1}{\milli\meter} as opposed to \SI{0.1}{\milli\meter}) flexures; however, the flexure length in such a resonator would be much larger for the same resonant frequency. For instance, COMSOL simulations suggest that \SI{1}{\milli\meter} thick flexures would need to be >\SI{0.5}{\meter} long to retain a resonance of \SI{3.76}{\hertz}. Such long flexures do not follow our development goals of compact and portable optomechanical inertial sensors. 

\section{Test Mass Displacement Interferometer}
\label{sec:displacement_interferometry}
\begin{figure}[htbp]
\centering
\fbox{\includegraphics[width=.95\linewidth]{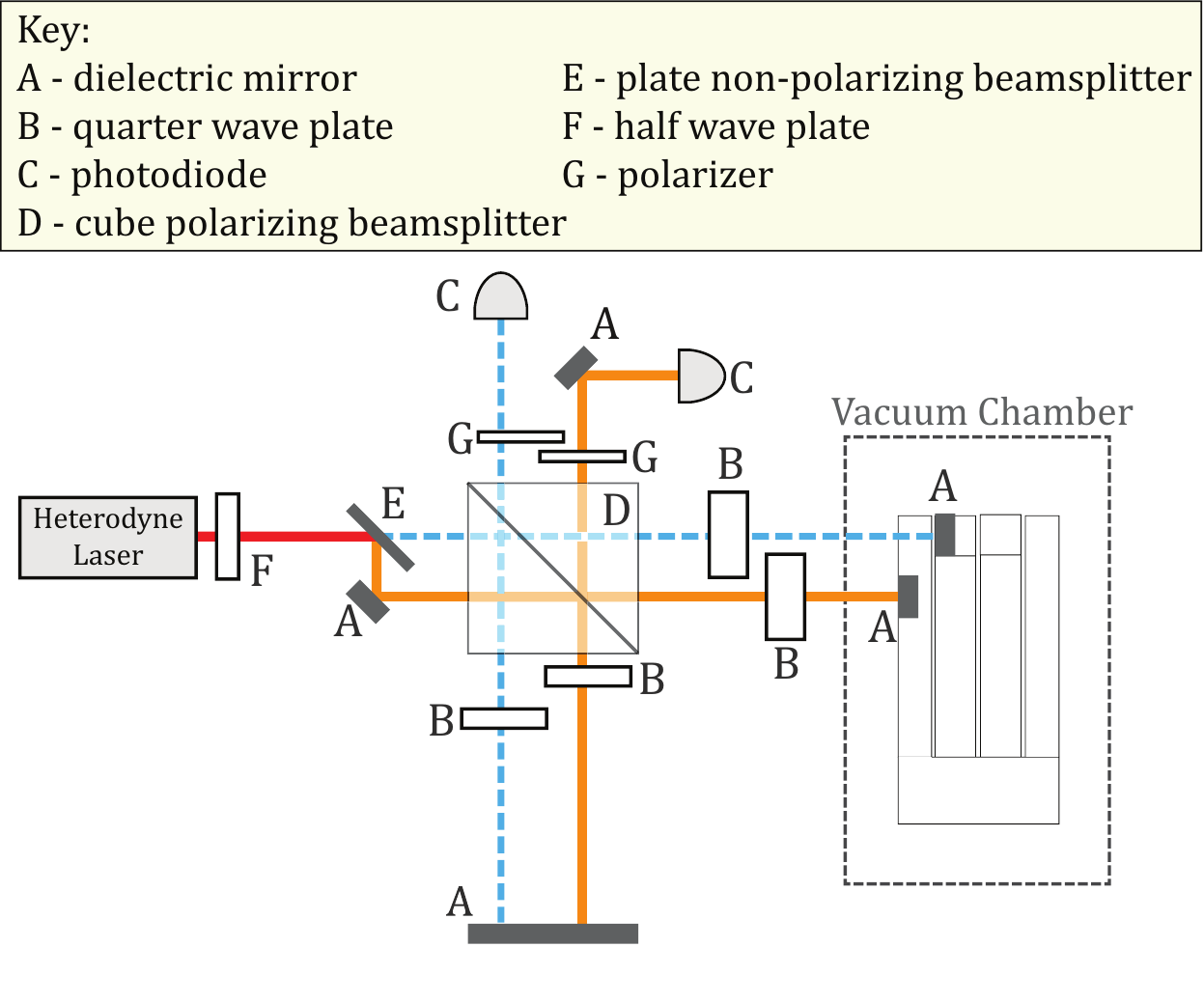}}
\caption{Diagram of the interferometers used to measure acceleration and displacement power spectral densities. A heterodyne laser beam consisting of two frequencies is split in two by a non-polarizing beam splitter. The light is equally split between the two interferometers, measurement and reference. The signal arm of measurement interferometer reflects off of the mirror placed on the test mass. The signal arm of the reference arm reflects off a mirror placed on the frame of the resonator. The reference arm of both interferometers reflect off a common mirror. The displacement of the test mass is measured by subtracting the phases of the two interferometers. Using two interferometers allows for the rejection of common-mode noise, lowering the total read-out noise.}
\label{fig:interferometer}
\end{figure}

In order to verify these models, we need a method for detecting test mass displacement. In this section, we outline the construction of a test read-out system for this purpose. However, this interferometer is used to characterize the resonance and quality factor of the mechanical resonator and is not developed to achieve high sensitivities. In the future, the resonator will be fully integrated with a high-sensitivity displacement readout interferometer~\cite{LaserInterferometry} to create the final optomechanical sensor. When conducting measurements on our sensor in a laboratory environment, we can expect test mass displacements well over several microns. We therefore require an interferometric readout method that provides a sufficiently large dynamic range and allows for high displacement sensitivity in future developments. To this end, we built a heterodyne laser interferometer (Figure~\ref{fig:interferometer}), which is capable of measuring displacements significantly larger than an interferometer fringe to characterize the mechanical resonator and directly measure its resonant frequency and $Q$. 

To track test mass displacement, we placed an external mirror on top of the test mass and mounted a second mirror on the frame of the resonator to provide an interferometric reference phase that allows for differential measurements. Except for the mirrors reflecting their respective signal arms, the two interferometers share many optical components to facilitate common-mode noise rejection.  To reduce gas damping, we placed the optomechanical resonator into a low-pressure chamber that reaches \SI{0.9}{\milli\torr}. This chamber contains a viewport for optical access to the two mirrors placed on the resonator. This interferometer is operated in air outside the chamber, except for the mirrors placed onto the resonator.

\section{Experimental results}
From the interferometers described above, we were able to perform preliminary tests of the COMSOL models. We measured the quality factor of the resonator from ringdown measurements, which analyze the decay envelope of the maximum test mass displacement over time. 

\begin{figure}[htbp]
\centering
\fbox{\includegraphics[width=.95\linewidth]{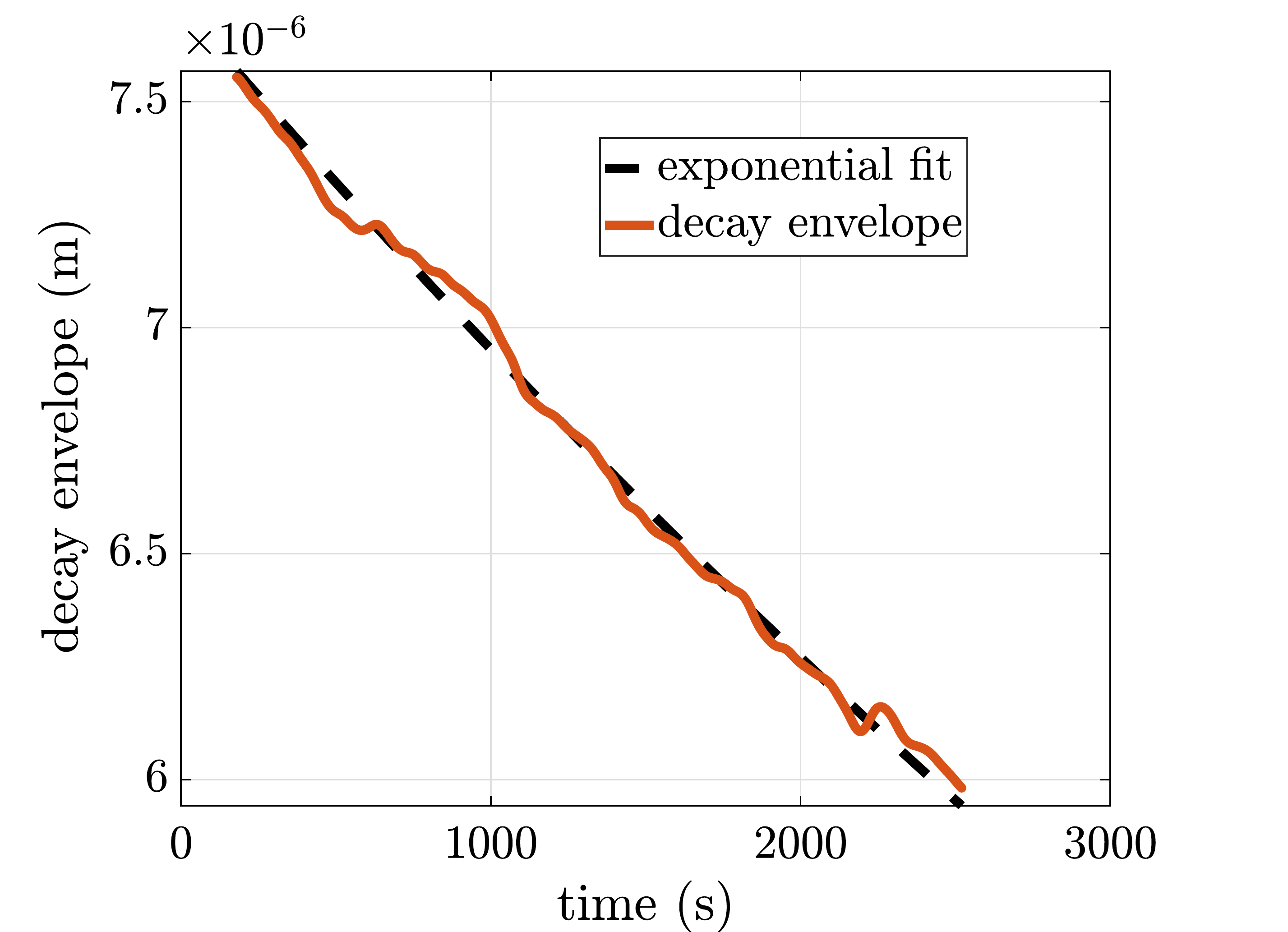}}
\caption{Decay envelope of the test mass oscillations during a ringdown. Fitting to an exponential decay, we find $Q=\num{1.14e5}$. This fit has an $r^{2}$ value of 0.989.}
\label{fig:envelope}
\end{figure}

Ringdown measurements at atmospheric pressure yielded quality factors of $Q = 600-700$, in good agreement with the COMSOL simulations. We then pumped the vacuum chamber down to \SI{0.9}{\milli\torr} and recorded a ringdown of the resonator over one hour. The decay envelope, shown in Figure~\ref{fig:envelope}, yields a mechanical quality factor of $Q=\num{1.14e5}$. This corresponds to an $mQ$-product of ~\SI{250}{\kilo\gram} and a thermal noise floor of $a_\textnormal{th} =\SI{4.03e-11}{\meter\per\second\squared\per\sqrt{\hertz}}$ at higher frequencies, increasing with a slope of $1/f$ towards low frequencies.  

\begin{figure}[htbp]
\centering
\fbox{\includegraphics[width=.95\linewidth]{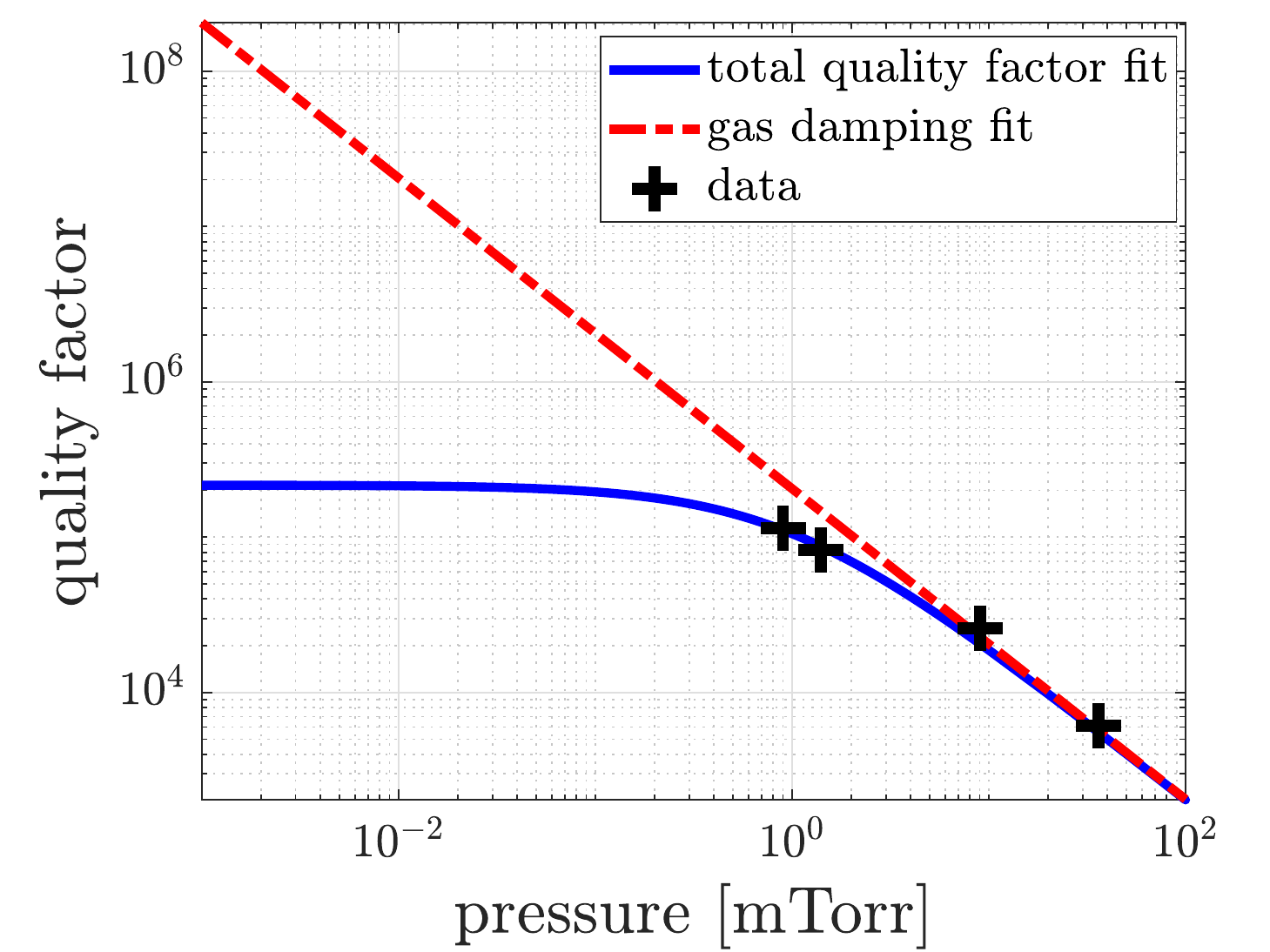}}
\caption{Quality factors obtained from ring-downs are plotted versus the pressure in the vacuum chamber. From the fit, we infer that the quality factor of the resonator is limited by gas damping at the pressures we can achieve.}
\label{fig:pressure}
\end{figure}

When studying and measuring the quality factor, we observed that there is a strong dependence on pressure, even when pumping down to the mTorr regime, which is expected. This behavior suggests that $Q$  is still limited by gas damping. Figure~\ref{fig:pressure} shows the quality factors we have observed versus pressure. Gas damping losses limit the quality factor of the resonator at this pressure regime to a so-called ballistic regime that can be determined by \cite{Bianco2006}:
\begin{equation}
    Q_{gas} = \frac{a}{P}
    \label{gas_damp_Q},
\end{equation}
where $a$ is a parameter dependent on temperature and the properties of the gas. The reciprocals of quality factors add linearly:
\begin{equation}
    \frac{1}{Q}=\frac{1}{Q_{gas}}+\frac{1}{Q_{0}},
\end{equation}
where $Q_{0}$ is the quality factor due to other loss mechanisms. By fitting the data in Figure~\ref{fig:pressure} we obtain that $a = \SI{205}{\torr}$ and $Q_{0} = \num{2.17e5}$. The $Q$ vs. pressure data points on the plot clearly follow a trend in agreement with gas damping, indicating that this the current dominant loss mechanism in our system for pressures at the \SI{}{mTorr} level.

\section{Outlook}
In this work, we modeled the energy loss mechanisms limiting the sensitivity of a novel \SI{10}{\hertz} optomechanical inertial sensor. In contrast to previous work in optomechanical accelerometers over kHz frequencies \cite{Guzman2014}, the low-frequency resonance of this sensor allows for better sensing of low-frequency signals. Using two heterodyne interferometers as displacement sensors, we presented preliminary measurements of the resonator's quality factor at various pressure levels. Improvements to our low-pressure and vacuum facilities are currently under commissioning, and we expect that these improvements will lead to significantly lower mechanical losses, laser-interferometric displacement and acceleration noise floors in our  optomechanical inertial sensors. We have demonstrated that our mechanical resonator can achieve a $mQ$-product of \SI{250}{\kilo\gram} under our current experiment conditions, leading to acceleration noise floors at levels of $\SI{1e-11}{\meter\per\second\squared\per\sqrt{\hertz}}$. However, we anticipate to achieve higher $mQ$-products as we improve our vacuum systems.

These investigations show that our sensor's compact dimensions, magnetic field insensitivity, and high mechanical quality factor make low-frequency optomechanical inertial sensors promising candidates for field high sensitivity acceleration measurements. 

\section{Funding Information}
This work was supported, in part, by the National Geospatial-Intelligence Agency (Grant Number: HMA04761912009) and the National Science Foundation (Grant Number: PHY-1912106).

\section{Acknowledgments}
The authors thank Cristina Guzman for reviewing and improving parts of the manuscript.

\bibliography{Reference}
 
\end{document}